\documentclass[manuscript]{aastex}

\shorttitle{LRLL54361}
\shortauthors{Balog et al.}

\begin{document}

\title{The extraordinary far-infrared variation of a protostar: \textit{Herschel}/PACS observations of LRLL54361\footnote{Herschel is an ESA space observatory with science instruments provided by European-led Principal Investigator consortia and with important participation from NASA.}}

\author{Zoltan Balog}
\affil{Max Planck Institute for Astronomy K\"onigstuhl 17, Heidelberg, D-69117, Germany}
\email{balog@mpia.de}

\author{James Muzerolle}
\affil{Space Telescope Science Institute, 3700 San Martin Dr., Baltimore, MD 21218, USA}

\author{Kevin Flaherty}
\affil{Astronomy Department, Wesleyan University, Middletown, CT 06459, USA}

\author{\"Ors H. Detre and Jeroen Bouwmann}
\affil{Max Planck Institute for Astronomy K\"onigstuhl 17, Heidelberg, D-69117, Germany}

\author{Elise Furlan\altaffilmark{*}}
\affil{Natinal Optical Astronomy Observatory, 950 N. Cherry Ave.,Tucson, AZ 85719, USA}
\altaffiltext{*}{Infrared Processing and Analysis Center, California Institute of Technology, 
770 S. Wilson Ave., Pasadena, CA 91125, USA }

\author{Rob Gutermuth}
\affil{Dept. of Astronomy, University of Massachusetts, Amherst, MA 01003, USA}

\author{Attila Juhasz}
\affil{Leiden Observatory, Leiden University, Niels Bohrweg 2, NL-2333-CA Leiden, The Netherlands}

\author{John Bally}
\affil{CASA, University of Colorado, CB 389, Boulder, CO 80309, USA}

\author{Markus Nielbock, Ulrich Klaas, Oliver Krause and Thomas Henning}
\affil{Max Planck Institute for Astronomy K\"onigstuhl 17, Heidelberg, D-69117, Germany}

\and

\author{Gabor Marton}
\affil{Konkoly Observatory, Research Center for Astronomy and Earth Sciences, Hungarian Academy of Sciences, Konkoly Thege 15-17, 1121 Budapest, Hungary}

\begin{abstract}
We report \textit{Herschel}/PACS photometric observations at 70 $\mu$m and 160 $\mu$m of LRLL54361 - a suspected binary protostar that exhibits periodic (P=25.34 days) flux variations at shorter wavelengths (3.6 $\mu$m and 4.5 $\mu$m) thought to be due to pulsed accretion caused by binary motion. The PACS observations show unprecedented flux variation at these far-infrared wavelengths that are well correlated with the variations at shorter wavelengths. At 70 $\mu$m the object increases its flux by a factor of six while at 160$\mu$m the change is about a factor of two, consistent with the wavelength dependence seen in the far-infrared spectra. The source is marginally resolved at 70 $\mu$m with varying FWHM. Deconvolved images of the sources show elongations exactly matching the outflow cavities traced by the scattered light observations. The spatial variations are anti-correlated with the flux variation indicating that a light echo is responsible for the changes in FWHM.  The observed far-infrared flux variability indicates that the disk and envelope of this source is periodically heated by the accretion pulses of the central source, and suggests that such long wavelength variability in general may provide a reasonable proxy for accretion variations in protostars. 
\end{abstract}


\section{Introduction}
During the formation process, a star slowly evolves from a core deeply embedded in gas and dust to a fully revealed star, possibly surrounded by planets. This evolution takes place over a few million years; however, there is growing evidence that the process is in fact highly dynamic on timescales from centuries to as short as days. This is one possible solution for the luminosity problem \citep[see e.g.][]{Keny90,Dunham13}. Most young stellar objects show variability at some level in the optical and near-infrared wavelength regime due to cold stellar spots related to stellar activity, hot spots associated with accretion flows striking the surface of the star, or variable extinction along the line-of-sight \citep[see e.g.][]{Carp01,Cody14}. In recent years it was recognized that flux changes in the mid-infrared are also common \citep{Morales09, VanBoek10,Morales11,Flaherty13}. However strong and periodic flux variation in the far-infrared (FIR) has not yet been reported in the literature. The variations discovered so far are either not periodic \citep[][]{Harv98, Kosp07} or weak, with amplitudes around 10-20\% \citep{Bill12}.  At these wavelengths the radiation is dominated by thermal emission of the circumstellar material, which could in principle vary as a consequence of either its changing illumination/accretion heating or a physical restructuring of the material. Thus, by observing variability of protostars in the far-IR regime we gain insight into how the envelope is heated by the central source, and how it may respond to changes in its illumination. Our knowledge about the flux variation of   sources that are still embedded in their envelope material is very limited.

Our target LRLL54361 is a young protostar in the star forming region IC 348, at a distance of 320 pc. This unique source exhibits strong, periodic increases in luminosity at mid-infrared wavelengths, as measured from multi-epoch \textit{Spitzer} Space Telescope observations taken over a time span of several years (\cite{Muze12}, hereafter M13). The photometric light curve shows a remarkably stable periodicity with a period of 25.34 $\pm$ 0.01 days. M13 also obtained near-infrared \textit{HST}/WFC3 images that revealed a light echo propagating through scattered light cavities in the envelope. The overall characteristics of the flux changes in this source were best explained as a result of pulsed accretion due to an unseen companion star.  In this work, we analyze new multi-epoch Herschel observations of LRLL54361 in order to better characterize the nature of this object and its variability.

\section{Observations and data processing}
\subsection{\textit{Herschel}/PACS Photometry}
We observed a 14$\arcmin$x14$\arcmin$ area in IC348 with the Photodetector Array Camera and Spectrometer \citep[PACS;][]{Pogl10} onboard the \textit{Herschel} Space Observatory \citep{Pilb10} simultaneously at 70 and 160 $\mu$m 24 times in scan map mode. An additional 5 epochs were observed during a later pulse phase of LRLL54361 in all three PACS Photometer bands. We processed our PACS maps up to the Level 1 stage using the Herschel Interactive Processing Environment \citep[HIPE, V11.0 user release,][]{Ott10}. For the final Level 2 maps and the combination of the different scan angles we used the Scanamorphos software \citep{Rous12}, which handles the ambient extended emission and the compact sources in our maps equally well. We used a uniform pixel scale in all of our maps (1$\arcsec$/pixel at 70 $\mu$m and 2$\arcsec$/pixel at 160 $\mu$m). 

Although most of IC348 is affected by strong diffuse background emission at long wavelengths, LRLL54361 lies in a relatively smooth area. That is why simple aperture photometry is sufficient to extract the source flux from the PACS maps.  The photometry was performed on the final map in each epoch using the annularSkyAperturePhotometry task in HIPE \citep[V11.0 user release,][]{Ott10}. We used a 10$\arcsec$ aperture and a sky annulus between 25$\arcsec$ and 35$\arcsec$. A sky annulus fairly distant to LRLL54361 was needed because it has a close companion (LRLL54362) that would influence the background determination in the case of annuli closer to the star. We accounted for the lost flux due to the limited size of our aperture through aperture correction (Balog et al. 2014) using the 
photApertureCorrectionPointSource task in HIPE. Note that we used exactly the same aperture and exactly the same annulus for each map hence aperture correction has no effect on the relative fluxes of LRLL54361.  The results of the photometry are summarized in Table \ref{tab:phot}.

As a sanity check, we have also used the boloSource() algorithm, which is a tool for separating the signal of the sources from the diffuse background in the timeline of the PACS photometer measurements \citep{Marton13}. This method confirms our aperture photometry results.
\clearpage
\begin{deluxetable}{llcccccccc}
\tablecolumns{10}
\tabletypesize{\scriptsize}
\tablewidth{0pc}
\tablecaption{Observing log of the observations and measured fluxes\tablenotemark{a}}
\tablehead{
\colhead{Cycle} & \colhead{Date} & \colhead{Julian Date} & \colhead{Phase\tablenotemark{b}} & \colhead{Flux70} & \colhead{Err70} & \colhead{Flux100} & \colhead{Err100} & \colhead{Flux160} & \colhead{Err160}\\
\colhead{} &\colhead{} & \colhead{} & \colhead{} & \colhead{[Jy]} & \colhead{[Jy]} &\colhead{[Jy]} & \colhead{[Jy]} & \colhead{[Jy]} & \colhead{[Jy]}}
\startdata
\cutinhead{Photometry}
1&2012-02-08 09:01:50.0 & 2455965.876 & 0.334 & 6.566 & 0.062 & \nodata & \nodata & 12.561 & 0.399\\ 
1&2012-02-09 07:59:10.0 & 2455966.833 & 0.371 & 5.748 & 0.062 & \nodata & \nodata & 12.026 & 0.388\\ 
1&2012-02-12 16:58:36.0 & 2455970.207 & 0.505 & 4.200 & 0.048 & \nodata & \nodata & 10.789 & 0.449\\
\cutinhead{Spectroscopy}
\nodata&2012-08-13 18:17:57 & 2456153.263 & 0.729& \nodata & \nodata & \nodata & \nodata & \nodata & \nodata\\
\nodata&2012-08-14 12:52:01 & 2456154.036 & 0.759& \nodata & \nodata & \nodata & \nodata & \nodata & \nodata\\
\nodata&2012-08-21 02:43:51 & 2456160.614 & 0.019& \nodata & \nodata & \nodata & \nodata & \nodata & \nodata\\
\nodata&2012-08-21 11:34:47 & 2456160.983 & 0.033& \nodata & \nodata & \nodata & \nodata & \nodata & \nodata
\enddata
\tablenotetext{a}{The full table is available in electronic form. Only a portion of it is shown here for guidance regarding its form and content.}
\tablenotetext{b}{We calculated the phase of the observations based on the period of 25.34 days and zero point JD 2455121.203 reported by M13. Phase=0 corresponds to the peak flux, as measured from the Spitzer/IRAC observations.}
\label{tab:phot}
\end{deluxetable}
\clearpage

\subsection{\textit{Herschel}/PACS Range Spectroscopy}
The PACS spectrograph consists of a 5$\times$5 array of 9.4'' $\times$9.4''
spatial pixels (hereafter referred to as spaxels) covering the
spectral range from 52-210~$\mu$m with $\lambda/\delta\lambda \sim $1000-3000.
Spectra were obtained in two spectral orders simultaneously, with the second order ranging from 51 - 105~$\mu$m
        and the first order from 102 - 210~$\mu$m.
The spatial resolution of PACS-S ranges from $\sim$9'' at 50~$\mu$m to $\sim$18'' at 210~$\mu$m.
Our target was observed in the standard range-scan spectroscopy mode
with a grating step size corresponding to Nyquist sampling (see
further \citet{Pogl10}). We obtained two spectra between 55-72 and 100-145~$\mu$m close to the minimum (at Phase=0.729 and 0.759) and two spectra 7 days later on as close to the predicted maximum as possible (at Phase=0.019 and 0.033). On both occasions the two observations were separated by several hours to study possible short term variations (see the observing log of the spectroscopic observation at the bottom of Table \ref{tab:phot}).

We processed our data using HIPE with calibration version 45 and standard pipeline scripts. 
The infrared background emission was removed using two chop-nod positions 6' from
the source in opposite directions. 
Absolute flux calibration was achieved by normalizing the source spectra to the emission spectrum of the telescope mirror 
itself as measured by the off-source positions, and by applying a detailed model of the telescope emission available in HIPE. 
Before combining the two nod positions and multiple grating scans, spectral rebinning was done with 
an oversampling of a factor of two and an upscaling of a factor of one corresponding to Nyquist sampling.  

Our target was well centered on the central spaxel which we
used to extract the spectra as this provided the highest signal-to-noise ratio (SNR) in the spectra. 
We detected LRLL54362 within the field of view of the spectrograph, centered on spaxel 14 at the edge of the image slicer.  We checked the absolute flux calibration of our spectra against the PACS photometry
taken close in time and found that the photometry matched the observed flux levels of our spectra very well. We expect, therefore, no substantial systematics in our flux calibration due to either telescope mispointing or from the neighboring source.

\section{Analysis and Results}
\subsection{Photometry}
\begin{figure}[h]
\epsscale{0.8}
\plotone{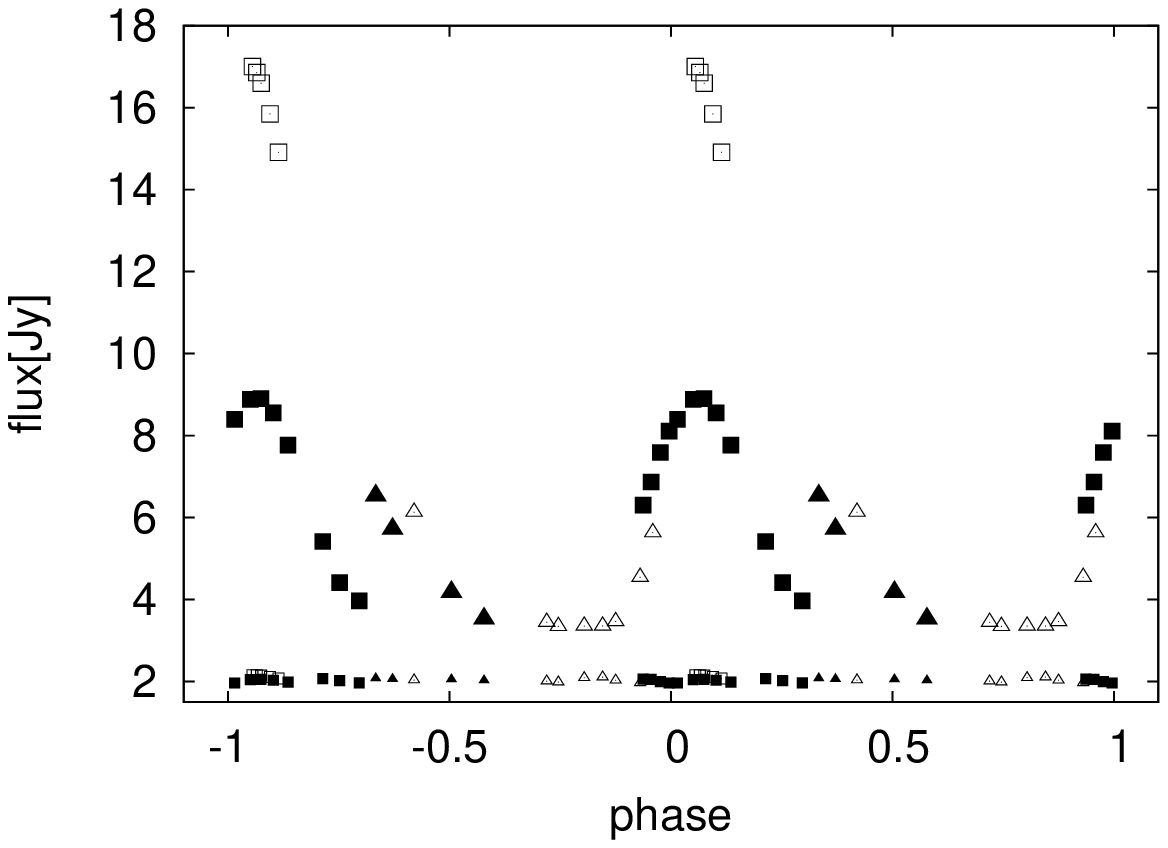}
\plotone{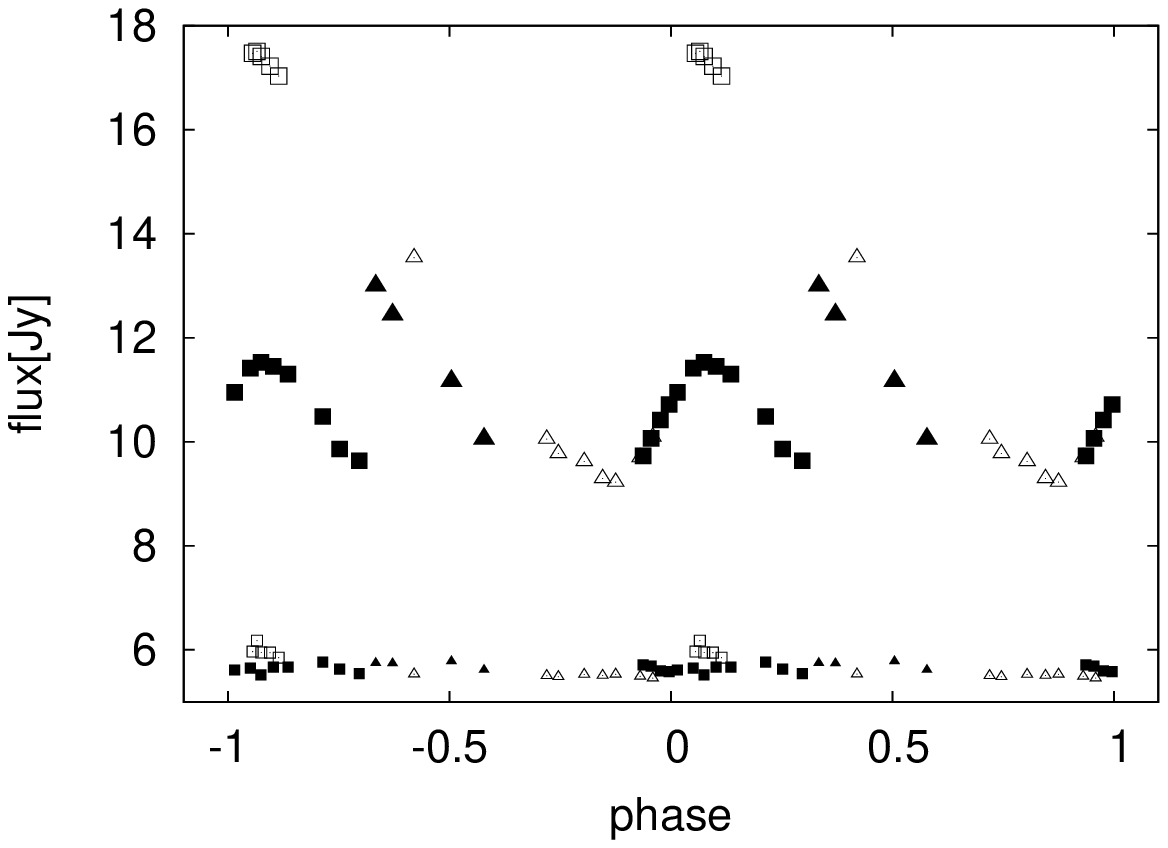}
\caption{Light curve of LRLL54361 folded with 25.34d period. Upper panel: 70 $\mu$m; lower panel 160 $\mu$m. Different symbols designate observations taken in different cycles (cycle 1: filled triangles, cycle 2: open triangles, cycle 3: filled squares, cycle 4: open squares). As a comparison the lightcurve of LRLL54362 (around 2Jy in the blue and 5.6Jy in the red) is also shown to demonstrate the stability of the instrument. The scatter of the flux of LRLL54362 is around 2\% in both bands.}
\label{fig:lc}
\end{figure}

Fig \ref{fig:lc} shows the folded light curve at 70 $\mu$m (upper panel) and 160$\mu$m (lower panel) of LRLL54361, assuming the 25.34-day period measured by M13 from Spitzer observations. The light curve at 100 $\mu$m is not shown because it covers only 5 epochs. The different symbols represent different observing campaigns (thus different cycles).
Our observations cover two flux maxima and a separate flux minimum. The change between the minimum and maximum flux at 70 $\mu$m is about a factor of 6 while at 160 $\mu$m it is about a factor of 2. The lightcurve of the neighboring star LRLL54362 is shown for comparison. It is also a protostar, but exhibits no variability in the mid-IR. From the light curve of LRLL54632 we calculate that our data processing and photometry are stable within about 2\% in accordance with the repeatability of constant flux sources \citep[see.][for details]{Balog13}.  We note that the flux of LRLL54362 is also slightly ($\sim$5\%) higher during the strongest pulse peak of LRLL54361. This elevated flux level is due to the fact that the two sources are very close to each other (about 20") and some flux from the wings of the point spread function (PSF) of LRLL54361 is present within the aperture of LRLL54362. The level of contamination calculated using the Encircled Energy Fractions \citep[Table 2]{Balog13} is about 0.5\% at 70 $\mu$m and 1\% at 160 $\mu$m. During the strongest pulse peak it amounts to $\sim$0.1 Jy at 70 $\mu$m and $\sim$0.2 Jy at 160 $\mu$m within the aperture of LRLL54362 which is consistent with the measured fluxes. Of course the flux of LRLL54362 also contaminates the flux of LRLL54361 but the contamination is well below 1\% of the flux of LRLL54361 even when it is at its faintest stage.  Since LRLL54362 exhibits no obvious flux changes, other than what can be ascribed to contamination from LRLL54361 and the contamination is very low, it has has no bearing on our results.

The Herschel flux variations of LRLL54361 are consistent with the 25.34-day period measured by M13. However, the variation pattern does not repeat exactly; Figure \ref{fig:lc} shows that the second maximum (open squares) is about 60-80\% brighter (depending on the filter) than the first one (filled squares).
There is a slight shift between the curves of the different cycles implying that the strength and/or the exact time of the pulse changes from cycle to cycle. The light curve shapes also imply that the system returns to the same minimum state after each pulse but the timescale of the decline depends on the amplitude of the maximum - the stronger the maximum, the more time it takes to return to the minimum. Quantitatively the decline at 70 $\mu$m can be described by an exponential curve with an $\sim$ 4.5 days time constant.
At 160$\mu$m the decline can also be described with an exponential function but the time constant is about an order of magnitude higher and so the decay time is considerably larger than one period.  
We note that there is a time-lag between the \textit{Spitzer}/IRAC and \textit{Herschel}/PACS data as well as between the PACS 70 $\mu$m and PACS 160 $\mu$m light curves. The maximum at 70 $\mu$m is delayed by 1.6 days with respect to the predicted maximum based on the Spitzer ephemeris. This cannot be considered as an absolute value since there is some phase jitter from cycle to cycle. 
However the maxima of the contemporary measurements at 70 and 160 $\mu$m differ by 0.7 days indicating that the 160$\mu$m emission region must be farther away from the star than that emitting at 70 $\mu$m, as expected from models of infalling envelopes \citep{Keny93} 

\subsection{Spectroscopy}
\begin{figure}[h]
\epsscale{1.0}
\plotone{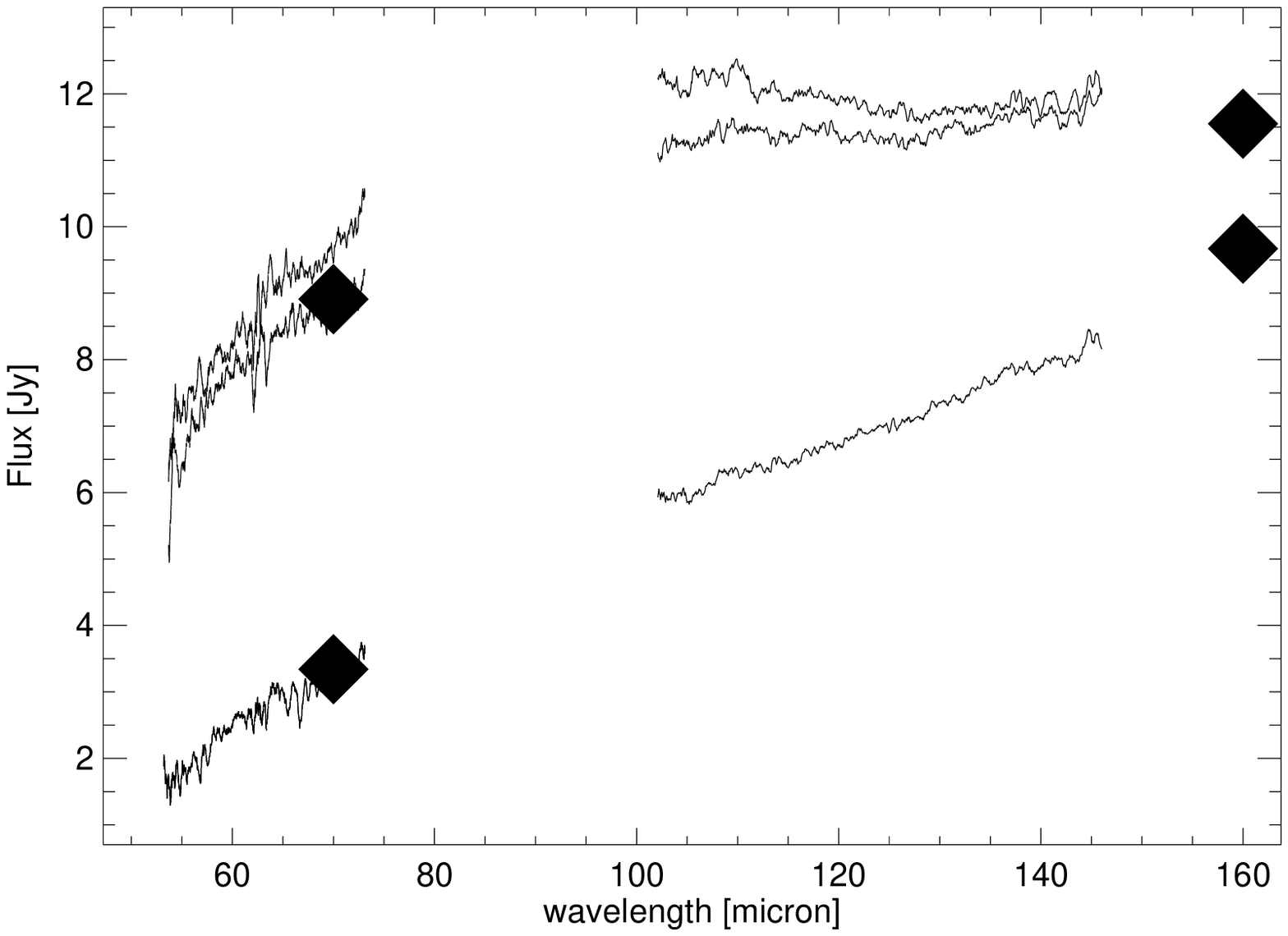}
\caption{HERSCHEL/PACS range spectroscopy. The lowermost spectrum is the average of the two spectra taken during the minimum phase. The upper two were taken at maximum phase. Diamonds show the PACS photometry values during minimum and maximum.}
\label{fig:spect}
\end{figure}

We show the results of the \textit{Herschel}/PACS range spectroscopy in Fig. \ref{fig:spect}. Since the two spectra taken at the minimum show no significant difference we averaged them to compensate for the lower signal to noise of these measurements. The two spectra taken at the maximum clearly show different continuum levels indicating that the object was still increasing in brightness during the two spectroscopic observations. We also plot the minimum and the maximum values of the PACS photometry in Fig. \ref{fig:spect}. We note that the PACS/Photometer measurements taken at maximum light are contemporary with the spectra (less than 2 hours difference between the photometry and the first spectrum) while the ones taken in the minimum are not. The agreement between the fluxes of the photometry and spectroscopy is excellent indicating that the flux calibration of the spectra is very reliable. These spectra show that the flux variations are larger at shorter wavelengths, again consistent with the photometry. There are no spectral features of significance in the data.

\subsection{Source morphology}

\begin{figure}[h]
\epsscale{1.0}
\plotone{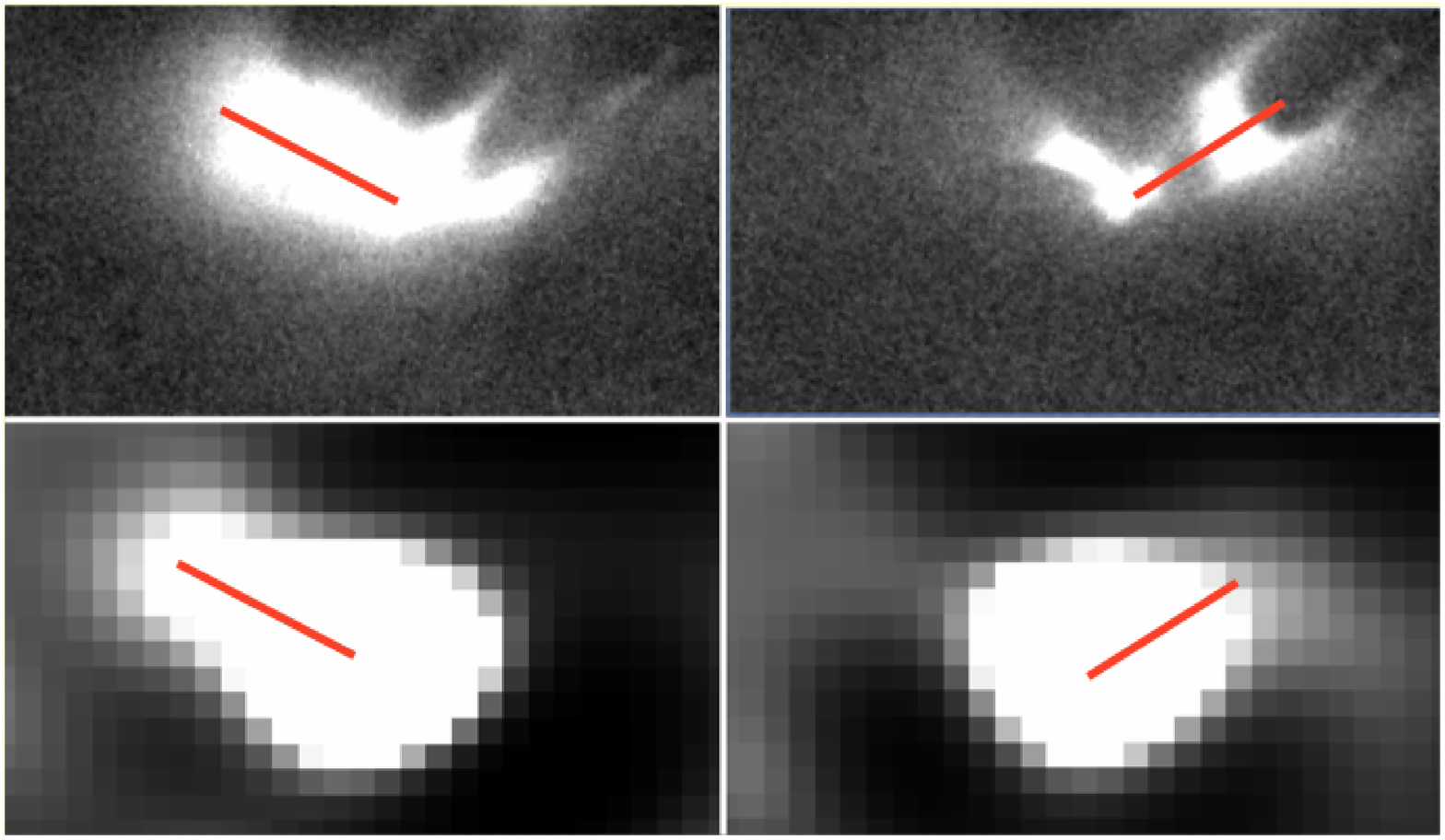}
\caption{Images of LRLL54361 from HST/WFC3-IR at 1.6 $/mu$m (top panels) and HERSCHEL/PACS at 70 $\mu$m (deconvolved with the PACS PSF to recover the spatial structure). The left panels show LRLL54361 8 days after maximum light, while the right panels show the source at its minimum. The red lines connect several photometric regions from the HST images measured by M13 (see their Fig. S4): regions A-F in the left panel and A-D in the right panel.}
\label{fig:image}
\end{figure}

During the 29 epochs of observations the shape of the source changed with the period of the photometric changes. Figure \ref{fig:image} compares deconvolved images of LRLL54361 at 70 $\mu$m to HST 1.6 $\mu$m images from M13 at two epochs, peak+8 days (corresponding to the largest spatial extent) and peak+18 days (when the central source is at flux minimum). In both cases an elongated structure appears to the side of the central source, indicating the light echo propagating outwards after the pulse reaches its peak. The appearance of the elongated 70 $\mu$m structure 8 days after the pulse peak is in excellent agreement with the timing and morphology of the HST images if we take into account the 1.6-day lag in the pulse timing between the Spitzer and Herschel wavelengths. The length of the elongated structures is about 2500 AU.  We interpret the far-IR light echo to represent a wave of dust heating in the outflow cavity walls. It is surprising to see such strong heating so far from the central source

To further study the morphology of the source at 70 $\mu$m we plotted the variation of the Full Width at Half Maximum (FWHM) of the source as a function of phase along the red line shown on the left panel of Fig \ref{fig:image}. To account for the PSF changes due to spacecraft rotation we divided the actual FWHM values by the FWHM of LRLL54362, the nearby non-variable source. In comparison to Fig. \ref{fig:lc} there is a clear anti-correlation between the size and brightness of LRLL54361, also indicating the far-IR light echo propagating outwards along the cavity wall. 

\begin{figure}[h]
\epsscale{1.0}
\plotone{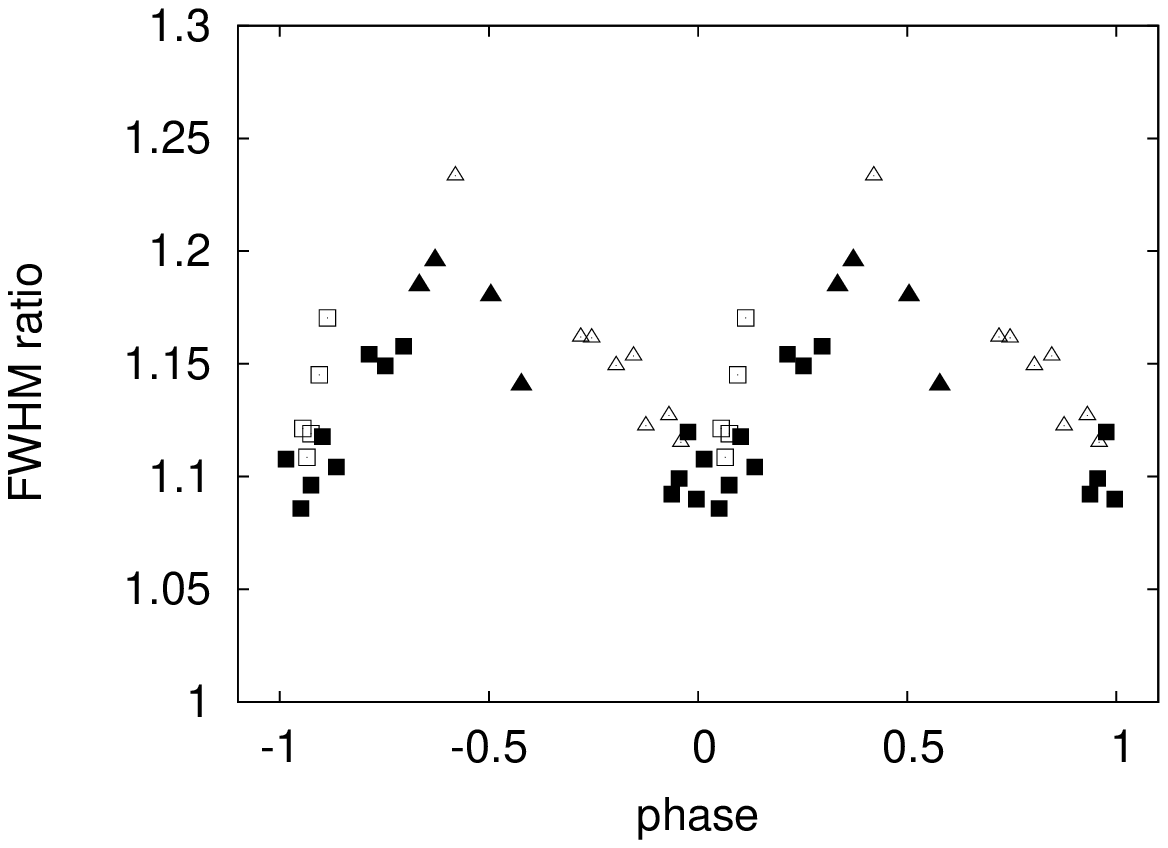}
\caption{The ratio of the FWHMs of LRLL54361 and LRLL54362 at 70 $\mu$m. Symbols as in Fig. \ref{fig:lc}.}
\label{fig:fwhm}
\end{figure}

\section{Discussion}

We have presented the first complete FIR light curve of the variable protostar LRLL54361. The light curves show large periodic variations at 70 and 160 $\mu$m (factor of 6 and 2, respectively) over a time span of $\sim 400$ days; this is also consistent with the wavelength-dependent variability seen in the PACS spectra. The FIR flux changes correlate very well with the period measured previously at shorter (3.6 and 4.5 $\mu$m) wavelengths, though with considerable differences in the peak flux level. The unresolved FIR emission traces cool dust several AU from the central source (and the resolved component is emitted even further away), thus the strength of its variability is somewhat surprising.  The periodicity and timescale automatically rule out a relation to \textit{in situ} structural variations because any such changes will occur on the local dynamical timescale, which for keplerian rotation in the disk is years or decades.

The most plausible mechanism is then variable irradiation of the disk and envelope either by means of time-dependent extinction of the central source or intrinsic variability of the source luminosity. The FIR fluctuations likely originate primarily from the inner few AU of the disk surface and the cavity walls in the infalling envelope, where the material is directly irradiated and diffusion timescales are not relevant. Both the wavelength dependence of the PACS spectra and the time lag between pulse peaks at different wavelengths are consistent with this picture, since shorter wavelengths have an increasing contribution from regions closer to the central object, which react faster to variations in the irradiation flux \citep{Jonhs13}. Variable extinction or obscuration has been invoked to explain strong optical and near-infrared variability in some young stellar objects, for example in the case of the precessing circumbinary disk around KH15D \citep{Herb10} and the protostar PTF 10nvg \citep{Hill13}.  However, the FIR variability in LRLL54361 would require an enormous change in extinction ($\Delta A_V \gtrsim 500$) every 25.34 days, and any obscuring material should lie in the disk plane and hence cannot explain the repeated light echo moving in a perpendicular direction through the outflow cavities. Thus, we favor intrinsic variability of the central source, most likely manifested via changes in the accretion luminosity.
M13 attributed the near- and mid-IR variability to pulsed accretion associated with a close binary.  In that scenario, the accretion luminosity varies according to the binary orbital phase, reaching a maximum near periastron separation.  The observed IR light indirectly traces that variability via irradiation of the circumstellar material.  The Herschel data are consistent with this hypothesis, though they do not directly constrain it. 

The strength of the pulses change from cycle to cycle indicating variations in accretion luminosity on a monthly timescale on top of the 25.34-day modulation. Assuming that the source luminosity at the pulse peaks is dominated by accretion luminosity, we can use our observations to estimate the short term variation of the mass accretion rate during the pulses. Using the protostellar models of M13, we extrapolated the bolometric luminosity from the observed FIR flux and adopted a stellar mass and radius of 0.2 $M_{\odot}$ and 2 $R{_\odot}$. Converting accretion luminosity to accretion rate via $L_{acc} \sim GM{_\odot}{\dot M}/R_{\odot}$, we then derive 8.0$\times10^{-7} M_{\odot}/yr$ and 1.5 $\times10^{-6} M_{\odot}/yr$ for the two maxima covered by our \textit{Herschel}/PACS observations (corresponding to $\sim$3$-$7$\times10^{-8} M_{\odot}$ accreted material per pulse). The mass accretion rate almost doubled from one maximum to the next meaning that the early phases of star formation cannot be in all cases properly described by models assuming steady or just slowly changing accretion rates. These results indicate that multi-epoch observations in the far-IR provide a useful probe of accretion variability in objects where direct measurements are infeasible.

\acknowledgments
We would like to thank the anonymous referee for the comments and suggestions that significantly improved the manuscript. We also would like to thank the entire Herschel Mission Planning group (especially Mark Kidger and Rosario Lorente) for scheduling our observations perfectly despite many constraints. Z.B., and M.N. are funded by Deutsches Zentrum f\"ur Luft- und Raumfahrt (DLR). RG gratefully acknowledges funding support from NASA ADAP grants NNX11AD14G and NNX13AF08G and Caltech/JPL awards 1373081, 1424329, and 1440160 in support of Spitzer Space Telescope observing programs.

\bibliographystyle{apj}

\end{document}